\documentclass[pre,aps,twocolumn,floatfix,superscriptaddress,showpacs,nofootinbib]{revtex4-1}
\usepackage{amsmath,amssymb}
\usepackage{float,graphicx}
\usepackage{subcaption}
\usepackage{color}
\usepackage{xcolor}

\providecommand{\U}[1]{\protect\rule{.1in}{.1in}}
\def\eqn {\begin{equation}}
\def\eeqn {\end{equation}}

\begin{document}

\title{The role of zero-clusters in exchange-driven growth with and without input}

\author{Emre Esenturk}
\email{E.esenturk.1@warwick.ac.uk}
\affiliation {Mathematics Institute, University of Warwick, Gibbet Hill Road, Coventry CV4 7AL, UK}

\author{Colm Connaughton}
\email{C.P.Connaughton@warwick.ac.uk} 
\affiliation {Mathematics Institute, University of Warwick, Gibbet Hill Road, Coventry CV4 7AL, UK}
\affiliation{Centre for Complexity Science, University of Warwick, Coventry CV4 7AL, UK}
\affiliation{London Mathematical Laboratory, 8 Margravine Gardens, London, W6 8RH, UK}

\date{\today}
\begin{abstract}
The exchange-driven growth model describes the mean field kinetics of a population of
composite particles (clusters) subject to pairwise exchange interactions. Exchange in this context means that upon interaction of two clusters, one loses a constituent unit (monomer) and the other gains this unit. Two variants of the exchange-driven  growth model appear in applications. They differ in whether clusters of zero size are considered active or passive. In the active case, clusters of size zero can acquire a monomer from clusters of positive size. In the passive case they cannot, meaning that clusters reaching size zero are effectively removed from the system. The large time behaviour is very different for the two variants of the model. We first consider an isolated system. In the passive case, the cluster size distribution tends towards a self-similar evolution and the typical cluster size  grows as a power of time. In the active case, we identify a broad class of kernels for which the the cluster size distribution tends to a non-trivial time-independent equilibrium in which the typical cluster size is finite. We next consider a non-isolated system in which monomers are input at a constant rate. In the passive case, the cluster size distribution again attains a self-similar profile in which the typical cluster size grows as a power of time. In the active case, a surprising new behavior is found: the cluster size distribution asymptotes to the same equilibrium profile found in the isolated case but with an amplitude that grows linearly in time.
  
\end{abstract}

\maketitle


\section{Introduction}

The kinetic theory of growth processes like aggregation and coarsening is used to model a diverse range of phenomena in the natural and social sciences \cite{wattis_introduction_2006,krapivsky_kinetic_2010}. Two of the most common models are the Smoluchowski model of binary aggregation \cite{leyvraz_scaling_2003} and the Becker-D\"{o}ring model of diffusion-driven coarsening \cite{penrose2001becker}. In the Smoluchowski model, clusters grow by binary interactions that result in merging of clusters. In the Becker-D\"{o}ring model, large clusters do not interact directly. Rather they emit and absorb elementary units of mass (monomers), a mechanism that allows some clusters to grow at the expense of others. The exchange-driven growth (EDG) model \cite{ben-naim_exchange-driven_2003} is intermediate between these two. It is similar to the Smoluchowski model in the sense that clusters interact directly through binary interactions.  Interactions result, not in mergers, but in the exchange of a single monomer. EDG is therefore similar to the Becker-D\"{o}ring model in the sense that clusters can only grow and shrink one monomer at a time.
The EDG model has been used in the socio-physics literature to model the dynamics of asset exchange \cite{ispolatov_wealth_1998} and migration between cities \cite{ke_kinetics_2002,lin_kinetics_2003}. More recently, it has received  attention from the non-equilibrium statistical mechanics and probability theory communities interested in coarsening dynamics  and condensation in interacting particle systems like the zero-range process \cite{groskinsky_condensation_2003}  and its generalisations \cite{chleboun_condensation_2014,chau_explosive_2015}. The common feature of these models is particles hopping between neighbouring sites on a lattice at rates that depend on the number of particles occupying the origin and destination sites \cite{godreche_coarsening_2016,beltran_martingale_2017}. The EDG model arises as the mean-field limit of such models \cite{jatuviriyapornchai_coarsening_2016} considering each lattice site to be a cluster and with the exchange interaction corresponding to a single particle hopping from one site to another.

There are two variants of the EDG model that differ in whether clusters of zero size are considered active or passive. In the passive case,  clusters reaching size zero cannot grow again by reacquiring a monomer from a cluster of positive size. This means they are effectively removed from the system. In the active case, clusters of size zero can reacquire a monomer from a cluster of positive size. Both variants are reasonable depending on modelling considerations. When considering asset exchange  for example \cite{ispolatov_wealth_1998}, clusters correspond to agents' wealth. Agents reaching zero wealth might be considered bankrupt. In this case, it is reasonable to treat zero clusters as passive. On the other hand, when considering interacting particle systems \cite{jatuviriyapornchai_coarsening_2016}, clusters of size zero simply correspond to empty sites. There is no a-priori reason to exclude models where particles can hop to empty sites. In this case, it is reasonable to treat zero clusters as active. While the distinction between the active and passive variants of the EDG model may seem minor, the large time behaviour is very different between the two. The main purpose of this paper is to categorise these differences for certain classes of kernel characterized by their asymptotic scaling properties. The aim is to eventually obtain a complete parametric understanding of the classes of asymptotic behavior possible in the EDG model as the kernel is varied. 

We consider isolated and non-isolated situations. By isolated we mean that the total mass in the system is fixed. In the non-isolated system, monomers are added to the system at a constant rate so that the total mass in the system grows linearly in time. One of the first parametric studies of EDG is the paper of Ben-Naim and Krapivsky \cite{ben-naim_exchange-driven_2003} which provides an extensive study of the properties of an isolated EDG model with passive zero clusters. This was extended to the non-isolated case with passive zero clusters in  \cite{krapivsky_mass_2015}. Consequently the original contributions of this work largely concern the case of active zero clusters although we provide, for the passive case, some new theoretical findings and present numerical simulations as benchmarks for completeness and comparison. 

The organisation of the paper is as follows. In section 2, we write down the mean-field EDG equations and define the scaling parameters of the model. Section 3 considers the isolated case. The evolution of the system when zero clusters are passive is compared against the evolution when zero clusters are active. Very distinct behaviour in the large time is found.  In the passive case, the cluster size distribution tends towards a self-similar evolution and the typical cluster size  grows as a power of time. By contrast, in the active case, we identify a broad class of kernels for which the the cluster size distribution tends to a non-trivial time-independent equilibrium in which the typical cluster size is finite. Section 4 considers the non-isolated case and contrasts the cases of passive and active zero clusters. Again the two variants of the model behave very differently. In the passive case, the cluster size distribution again attains a self-similar profile in which the typical cluster size grows as a power of time. In the active case, a surprising new behavior is found: the cluster size distribution asymptotes to the same equilibrium profile found in the isolated case but with an amplitude that grows linearly in time. Section 5 concludes the article.

\section{Definition and basic properties of the Exchange-driven Growth (EDG) model}

At the kinetic level of description, one is interested in knowing the evolution
of the cluster size distribution $c_{j}(t)$ (the proportion of clusters with size
$j),$ where $t$ is the time. Symbolically, the exchange process can be
described as
\[
<j>\oplus<k>\rightarrow<j\pm1>\oplus<k\mp1>
\]
where inside of the brackets denotes the sizes of the clusters. Supposing
spatial uniformity, the rate of interaction between clusters of sizes  $<j>$ and $<k>$
is given by $K(j,k)\, c_{j}\, c_{k}$.  The product $c_j\,c_k$ reflects the assumption that the system is well mixed so that the law of mass action applies. The kernel, $K(j, k)$, is the rate at which a cluster of size $j$ gives a 
monomer to a cluster of size $k$. The microscopic details of the the nature of the exchange interaction are encoded in this kernel. Unlike the Smoluchowski model of aggregation-driven
growth, there is no a-priori reason why $K(j,k)$ should be symmetric for the EDG model. Hence, throughout the paper we will allow both symmetric and non-symmetric forms for $K(j,k)$.

The evolution of the isolated EDG system (without any source or sink
terms) is given by%
\begin{equation}
\dot{c}_{0}=c_{1}\sum_{k=0}^{\infty}K(1,k)c_{k}-c_{0}\sum_{k=1}^{\infty
}K(k,0)c_{k}\text{ }\label{infode-0}%
\end{equation}

\begin{align}
\nonumber \text{ }\dot{c}_{j} &  =c_{j+1}\sum_{k=0}^{\infty}K(j+1,k)c_{k}-c_{j}%
\sum_{k=0}^{\infty}K(j,k)c_{k}\\
&  -c_{j}\sum_{k=1}^{\infty}K(k,j)c_{k}+c_{j-1}\sum_{k=1}^{\infty
}K(k,j-1)c_{k}\text{ \ } \label{infode-j}%
\end{align}
for $j=0,1,2,...$ with given initial conditions%
\begin{equation}
c_{j}(0)=c_{0j}.\label{infIC}%
\end{equation}
These equations correspond to the EDG model with active zero clusters. In the passive case,  zero clusters are not allowed to reacquire monomers. To arrive at the corresponding equations, we set $K(j,0)=0$ and disregard the equation for zero clusters (as it would decouple from the rest of the system), starting the sums from $k=1$ in all equations. 

 The main ingredient determining the evolution of the EDG system is the kernel. As in other well studied growth models (e.g. Smoluchowski), what matters most for the large-time evolution of the system, is the asymptotic dependence of the kernel on the cluster sizes. Motivated by interest in scaling properties in applications, $K(i,j)$ is taken to be a homogeneous function, or sometimes asymptotically homogeneous for large arguments. Heuristic arguments suggest that for the passive case, it is the degree of homogeneity of $K(i,j)$ and the way in which this degree is split between $i$ and $j$, that controls the long-time behavior. Moreover, for the active case, it will be seen that the dependence of the kernel on the sizes of the departure and arrival clusters has a big role in driving the system to non-trivial equilibrium. Hence, we will study EDG systems with kernels which are simple enough to allow detailed analysis but also general enough to capture the crucial differences in the behaviour of solutions. With this in mind we assume, for the passive problem, $K(j,k)=j^{\mu}k^{\nu}$ and perform a parametric study (of the exponents) on this system by varying the cluster dependence (and also the homogeneity) of the kernel. For the active zero clusters, as noted in Section I, the forms of the kernels should reflect the key function that that active-zero cluster play. Hence, we will appropriately assume the general form $K(j,k)=\frac{1}{2}j^{\mu}(1+k^{\nu})$ for the kernel. 
  
 Another noteworthy similarity of the EDG system to Smoluchoswki model is that, in the passive case, for sufficiently fast growing kernels, the growth mechanism generates clusters of infinite size (i.e., gelation) in finite time (or even instantly) where the mass conservation breaks down. Such singular behaviour occurs for when $\mu>1$ for non-symmetric kernels and when $\mu>3/2$ for symmetric kernels ($K(j,k)=j^{\mu}k^{\mu}$). In this article we focus on the non-gelling regime where the rigorous results provide the comfort of existence of global unique solutions \cite{esenturk_mathematical_2018}

\section{Exchange-driven growth in an isolated system}

This is the simplest scenario where one can observe the differences in the dynamics of EDG with active versus passive zero clusters.
In the case of passive zero clusters the growth is unidirectional and arbitrarily large sizes are generated while smaller clusters disappear in time \cite{ben-naim_exchange-driven_2003}. For the case of active zero clusters, transfer of monomers from finite-sized clusters to zero-clusters acts like a stabilization mechanism. This can impede the growth and one might expect that this could lead to equilibriation. We show here that that while this intuition  is correct for some kernels, the existence of a non-trivial equilibrium is not obvious in general. Furthemore, even when a non-trivial steady state exists, its stability is not trivial with examples of unstable steady states known in aggregation \cite{ball2012collective} and aggregation-fragmentation \cite{matveev2017oscillations} models. Here we demonstrate numerically that, for the kernels that admit equilibrium steady states in EDG, time dependent solutions approach to these equilibria in the limit of large time. Recently, some mathematically rigorous sufficient conditions for the existence of non-trivial equilibrium and its attractor property was provided in \cite{schlichting_exchange-driven_2018} and \cite{esenturk_exchange-driven_2019}, the latter with explicit rates of convergence, emphasizing, in particular, the dependence of the kernel on the sizes of departure and arrival clusters.


\subsection{Passive zero clusters: Continuous growth and self-similarity}
 
We begin by studying kernels in the generalized product form 
$K(j,k)=j^{\mu }k^{\nu }$ with varying exponents. In all of the computations and calculations below, we focus on the non-gelling regime in order to avoid the singularities. For the numerical computations, time dependent cluster densities are obtained by truncating the system of equations at a finite (maximal) cluster size ($1000<N<10000$) and numerically solving the resulting finite system with appropriate initial conditions.

We first consider non-symmetric kernels of the form $K(j,k)=j^{\mu }k^{\nu }\theta (k)$ (where $\theta$ is the step function taking the unit value for positive arguments and zero elsewhere). Figure 1 shows the plot of cluster sizes $c_{1},$ $c_{2},$ $c_{3}$ versus time for symmetric kernels with exponents $\mu=1, \nu=0$ and $\mu=0, \nu=1$ with monodisperse initial conditions. It is seen that, all $c_{j}$ values decay to zero, that is, larger and larger clusters are generated in time. For the case $\mu=0, \nu=1$ we report (for the first time to the best of our knowledge) the exact solutions which is given by 

\[
c_{j}(t)=\left(  \frac{t}{1+t}\right)  ^{j-1}\frac{1}{(1+t)^{2}}%
\]
One sees that the difference between the exact solutions and the numerical solutions is virtually zero (see the Appendices for the derivation), providing a benchmark test of the numerical integration.%

\begin{figure}[h!]%
\centering

\begin{subfigure}[b]{0.9\linewidth}
    \includegraphics[width=\linewidth]{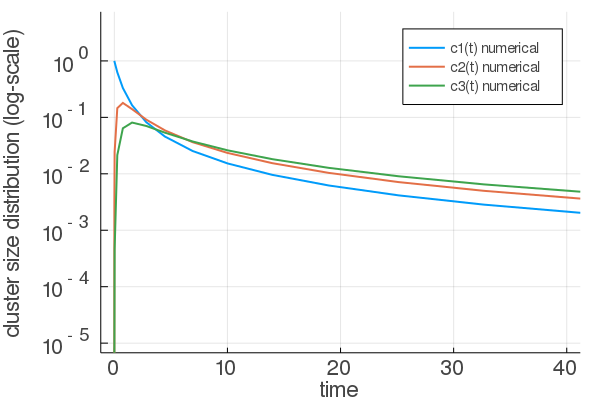}
    \caption{$\mu=1, \nu=0 $}
  \end{subfigure}

 \begin{subfigure}[b]{0.9\linewidth}
    \includegraphics[width=\linewidth]{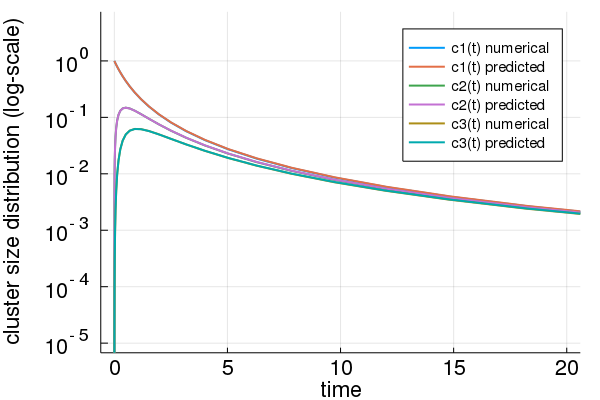}
    \caption{$\mu=0, \nu=1$}
    
  \end{subfigure}

\caption{Cluster densities (log-scale) for the isolated EDG system with passive
 zero clusters for non-symmetric kernels}%
\end{figure}

For symmetric kernels (of the form $K(j,k)=j^{\mu }k^{\mu }$) similar behaviour occurs, i.e., cluster densities decay in time as shown in Figure 2$.$ For the case $\mu=\nu=1$ analytic solutions \cite{ben-naim_exchange-driven_2003} are also plotted in the same figure.

\begin{figure}[h!]%
\centering

\begin{subfigure}[b]{0.9\linewidth}
    \includegraphics[width=\linewidth]{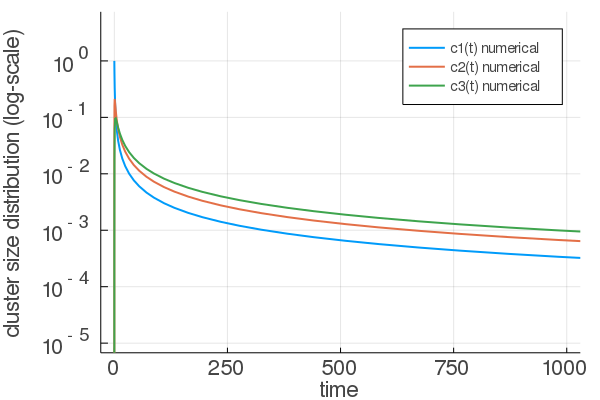}
    \caption{$\mu=0.0$}
  \end{subfigure}

  \begin{subfigure}[b]{0.9\linewidth}
    \includegraphics[width=\linewidth]{figs/class-c1c3-mu10-J00-teta00-g00-N4000-sR1-log.png}
    \caption{$\mu=1.0$}
  \end{subfigure}

 \begin{subfigure}[b]{0.9\linewidth}
    \includegraphics[width=\linewidth]{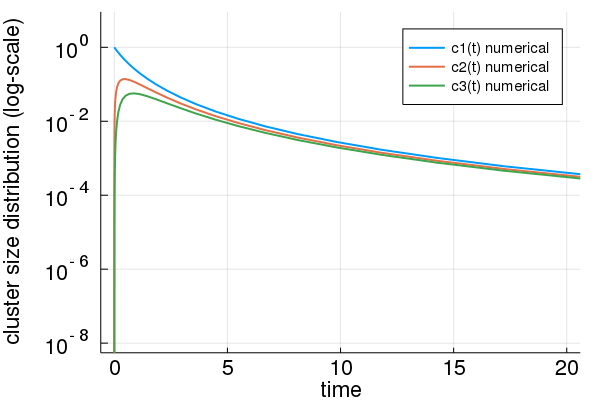}
    \caption{$\mu=1.2$}
  \end{subfigure}

\caption{Cluster densities (log-scale) for the isolated EDG system with passive
 zero clusters: $\mu=0.0$ (top); $\mu=1.0$ (middle); $\mu=1.2$ (bottom)}%
\end{figure}

Next, we numerically demonstrate the approach of solutions to self-similar forms. This was first shown in \cite{ben-naim_exchange-driven_2003} exploiting the similarity of EDG equations to the heat equation as we argue below. For symmetric kernels, $\mu=\nu $, 
by the separability of the kernels the evolution of the clusters are given by 
\begin{equation}
\text{ }\dot{c}_{j}=\left( (j+1)^{\mu }c_{j+1}-2j^{\mu }c_{j}+(j-1)^{\mu
}c_{j-1}\text{ }\right) \text{\ }(\sum_{k=1}^{\infty }k^{\mu }c_{k})
\end{equation}%

We know from the numerics that, over a period of time, most of the mass in the
system accumulates in large clusters and the unit mass exchange between
clusters can be regarded as incremental with respect to the cluster sizes.
Therefore in the large time limit the discrete difference can be approximated
by a second order derivative with respect to the cluster size. Then the 
continuum approximation of the infinite coupled ODE system is given by
\begin{equation}
\partial_{t}c(t,x)=\partial_{xx}(x^{\mu}c(t,x))\int_{x=0}^{\infty }x^{\mu }c(t,x),\label{PDE-iso}%
\end{equation}
It was shown in \cite{ben-naim_exchange-driven_2003} that this equation
admits scaling solutions where the typical cluster size grows as
$t^{\frac{1}{3-2\mu}}$ in time. Figure 3 demonstrates (for the
same set of kernels and the exponents) that the system evolves towards a
solution of the form $c_{j}(t)=\frac{1}{s(t)^{2}}\phi(\frac{j}{s(t)})$ with
$s(t)=t^{\frac{1}{3-2\mu}}$ as predicted. The scaling function is dependent on the exponent $\mu$
and, in principle, can be obtained by solving the reduced integro-differential equation.

\begin{figure}[h!]%
\centering

\begin{subfigure}[b]{0.9\linewidth}
    \includegraphics[width=\linewidth]{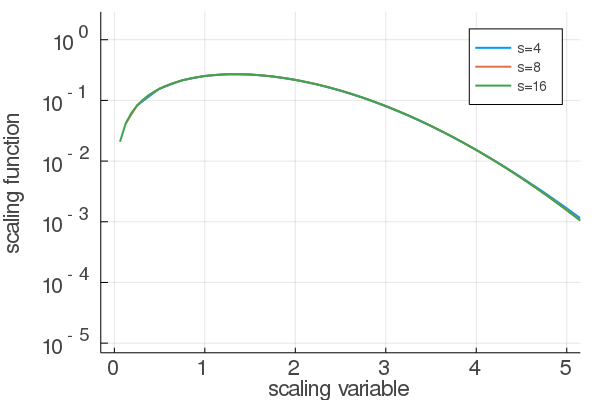}
    \caption{$\mu=0.0$}
  \end{subfigure}

  \begin{subfigure}[b]{0.9\linewidth}
    \includegraphics[width=\linewidth]{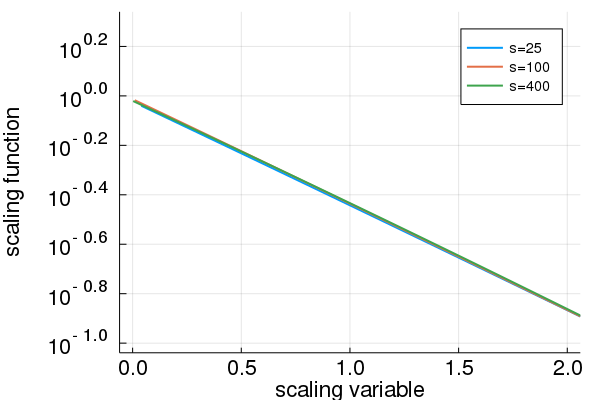}
    \caption{$\mu=1.0$}
  \end{subfigure}

 \begin{subfigure}[b]{0.9\linewidth}
    \includegraphics[width=\linewidth]{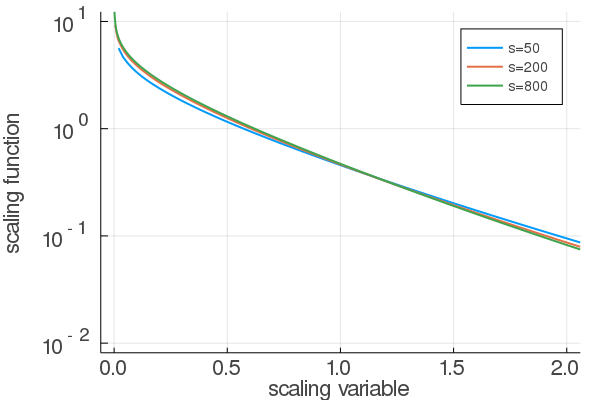}
    \caption{$\mu=1.2$}
  \end{subfigure}

\caption{Self similar Profile (log-scale) for the isolated EDG system with passive
 zero clusters: $\mu=0.0$ (top); $\mu=1.0$ (middle); $\mu=1.2$ (bottom)}%
\end{figure}

\subsection{Active zero-clusters and approach to equilibrium}

In the previous subsection, we saw that, for the case of passive zero clusters, the typical cluster size grows indefinitely and $\lambda-$moments go to infinity for $\lambda>1$. Here, we show that, for active zero clusters when monomers can be transferred to zero clusters, the story is quite different. 
We first discuss the existence of non-trivial equilibrium solutions where there is no growth at all. Non-trivial" here means that the cluster size distribution does not vanish. This excludes the possibility of all mass accumulating in a single large cluster which could be considered to be an equilibrium of sorts.The cases where there is no non-trivial equilibrium are identified. This is followed by presentation of specific exactly solvable cases where it is explicitly shown that the time dependent cluster densities approach to the values that correspond equilibrium densities of the same total mass and total number (of clusters). After this, using moment analysis and numerical computations we demonstrate that, for a range of exponents, all moments are bounded and time dependent solutions approach to the non-trivial equilibrium. The general convergence of solutions have
been shown rigorously by \cite{schlichting_exchange-driven_2018} for a large class of kernels and by one of the authors \cite{esenturk_exchange-driven_2019} for separable kernels of the forms
$K(j,k)=a_{j}b_{k}$ and $K(j,k)=ja_{k}+b_{j}+\varepsilon\beta_{j}\alpha_{k}$ with explicit rate of convergence. We remark that, neither the existence of non-trivial equilibrium solutions nor the convergence of solutions to the equilibrium follows necessarily from the inclusion of active zero clusters.

\subsubsection{\textbf{Existence and nonexistence of equilibria}}
{\color{red}

}
We first show that equilibrium solutions can
be constructed explicitly for a range of kernels which have the general separable form
\begin{equation}
    K(j,k) = a(j)\,b(k).
\end{equation}
Define
\begin{align*}
    A &= \sum_{k=1}^\infty a(k)\,c_k\\
    B &= \sum_{k=0}^\infty b(k)\,c_k\\
\end{align*}
and assume these sums are finite. Using Eq.~(\ref{infode-0}), we get
\begin{equation}
    c_1 = \frac{b(0)\,A}{a(1)\,B}\,c_0.
\end{equation}
Now iterating Eq.~(\ref{infode-j}), we find
\begin{equation}
\label{eqm-separable}
c_{j}=Q_{j}\left( \frac{B}{A}\right) ^{j}c_{0},\text{ \ \ \ }%
Q_{j}=\prod\limits_{k=1}^{j}\frac{b(k\!-\!1)}{a(k)}
\end{equation}
which can then be confirmed by induction. We require the total mass, $M_1 = \sum_{k=1}^\infty\,k\,c_k$, to be finite. Applying the ratio test to the sum $M_1$ using Eq.~(\ref{eqm-separable}) shows that there is an asymptotic restriction on the kernel:
\begin{equation}
\label{eqm-existence}
\lim_{j\to \infty}\frac{b(j)/j}{a(j+1)/(j+1)} < \infty.
\end{equation}
For the kernels we are interested in this article, $a(j)=j^\mu$ and $b(k) = 1 + k^\nu$, the condition Eq.~(\ref{eqm-existence}) implies that $\mu \geq \nu$ in order for the equilibrium to exist since the radius of convergence of the series $\sum_{j\geq1}jQ_{j}(\frac{B}{A})^j$ has to be finite. As a result we have the following 

(i) For $\mu > \nu$, non-trivial equilibrium exists always, that is, for all values of initial mass one can construct equilibrium solutions.

(ii) For $\mu < \nu$, there is no non-trivial equilibrium and all of the mass eventually concentrates at infinity

(iii) For $\mu = \nu$, there exists equilibrium up to a critical value of the initial mass. If the initial mass is greater than the critical value, there will not be an equilibrium and the excess mass will be concentrated at infinity.

It is worth noting that the equilibrium solutions
constructed here satisfy the detailed balance conditions. In other words, for each $j,k$ non-negative, the forward and backward rates in the exchange dynamics 
\[
<j>\oplus<k>\rightleftharpoons <j\pm1>\oplus<k\mp1>
\]
are equal i.e.,
\[
\sum_{k\geq 0}K(j,k)c_{k}c_{j}=\sum_{k\geq 0}K(k+1,j-1)c_{k+1}c_{j-1}
\]
which follows from the separability of the kernel and the recursive relationship between the clusters. 

Next, we show that the equilibrium solutions can be made more explicit for some integer exponents. Assume $M_{1}=M_{0}=1$ for simplicity (the case $M_{1},M_{0}\neq 1$ is
similar).  

Consider first the case for $\mu=\nu=0.$ Let $y=\sum_{j=1}^{\infty}c_{j}.$ Then
$c_{0}=1-y.$ Then by (\ref{eqm-separable}) one has
form
\[
c_{j}=y^{j}c_{0}
\]
Since $M_{1}=1,$ one has%
\[
1=\sum_{j\geq1}jy^{j}c_{0}.
\]
from which one obtains $y=\frac{1}{2}$ and hence $c_{0}=\frac{1}{2}.$ Therefore
one has
\begin{equation}
\label{equi-00}
c_{j}=\frac{1}{2^{j+1}}.
\end{equation}
Secondly, for the case $\mu=1, \nu=0$ the recursive relation yields $c_{j}=\frac{c_{0}}{j!}.$
By normalization $1=\sum_{j\geq0}c_{j},$ one finds $c_{0}=e^{-1}.$ Hence, the
cluster densities are given by
\begin{equation}
\label{equi-10}
c_{j}=\frac{e^{-1}}{j!}.
\end{equation}
For the integer exponents $\mu=1, \nu=1$ the combinatorial term $Q_{j}$ reduces to unity by cancellations and one finds the same cluster densities as in the constant kernel case.

Finally, we give a brief example of the non-existence which will be referred later. Let $\mu=0, \nu=1$. In this case $Q_{j}=j!$ and the series $\sum_{j\geq1}jQ_{j}z^j$ has zero radius of convergence which implies there is no admissible equilibrium. The interpretation is that when the kernel increases sufficiently fast with the size of destination cluster, the transfer to zero clusters cannot impede the overall mass transfer to larger clusters and the equilibrium cannot be sustained.

Next, we discuss the dynamics of the EDG problem.

\subsubsection{\textbf{Some exact solutions for the time-dependent EDG equations with active zero clusters}}

When the exponents in the kernel have integer values, it
is possible to construct explicit solutions to the infinite ODE system
(\ref{infode-0})-(\ref{infIC}). 
We first consider the kernel given by $\mu=1, \nu=0$ ($K(j,k)=j).$ In
this case, the rate equations read%
\[
\dot{c}_{0}=c_{1}M_{0}-c_{0}M_{1},
\]%
\[
\dot{c}_{j}=((j+1)c_{j+1}-jc_{j})M_{0}-(c_{j}-c_{j-1})M_{1}.
\]
where $M_{0},M_{1}$ are constant in time \cite{esenturk_mathematical_2018}. Using the generating function method (see the Appendices) one gets, for the cluster densities,
\[
c_{j}(t)=(M_{0}-M_{1}e^{-M_{0}t})e^{-g(t)}\frac{g(t)^{j}}{j!}+e^{-M_{0}%
t-g(t)}\frac{g(t)^{j-1}}{(j-1)!}.
\]
where $g(t)=(1-e^{-M_{0}t})M_{1}/M_{0}.$ We see that as $t$ goes to infinity cluster size distributions tend to constant values given by%
\begin{equation}
\label{eq-eqm}
c_{j}(\infty)=M_{0}e^{-M_{1}/M_{0}}\frac{(M_{1}/M_{0})^{j}}{j!},
\end{equation}
which is in accordance with the equilibrium solution, Eq.~(\ref{equi-10}), that we obtained in the previous section. 

Secondly, we vary the second exponent and consider $\mu=\nu=1$, that is, $K(j,k)=\frac{1}{2}j(1+k).$ In this case the first few cluster densities can be computed as \cite{mim_thesis}
\begin{equation*}
c_{0}(t)=\frac{2e^{t/2}(e^{t/2}-1)}{(2e^{t/2}-1)^{2}},\ \ c_{1}(t)=\ \frac{%
e^{t/2}(2e^{t}-3e^{t/2}+2)}{(2e^{t/2}-1)^{3}}
\end{equation*}%
\begin{equation*}
c_{2}(t)=\frac{2e^{t/2}(e^{t/2}-1)(e^{t}-e^{t/2}+1)}{(2e^{t/2}-1)^{4}},
\end{equation*}

\begin{equation*}
c_{3}(t)=\ \frac{e^{t/2}(2e^{2t}-5e^{3t/2}+6e^{t}-5e^{t/2}+2)}{(2e^{t/2}-1)^{5}}
\end{equation*}

\begin{figure}[h!]%
\centering

 \begin{subfigure}[b]{0.9\linewidth}
    \includegraphics[width=\linewidth]{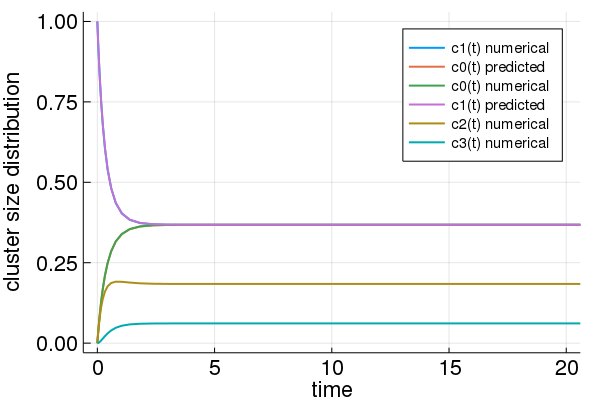}
    \caption{$\mu=1, \nu=0$}
  \end{subfigure}
  
  \begin{subfigure}[b]{0.9\linewidth}
    \includegraphics[width=\linewidth]{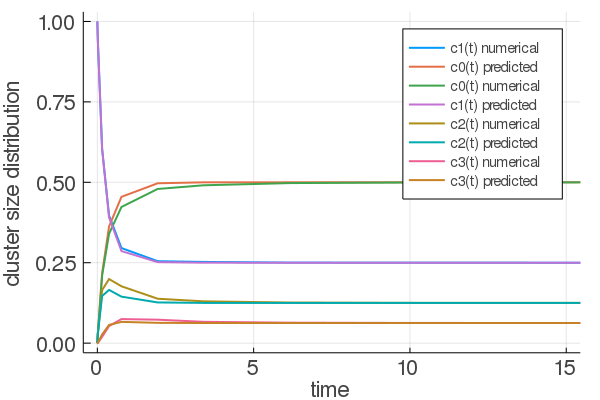}
    \caption{$\mu=1, \nu=1$}
  \end{subfigure}
  
  \begin{subfigure}[b]{0.9\linewidth}
    \includegraphics[width=\linewidth]{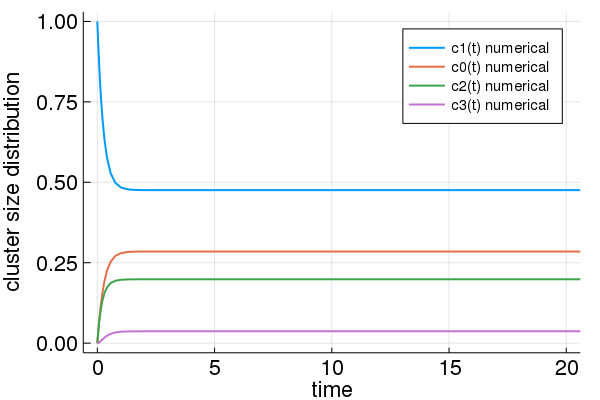}
    \caption{$\mu=2,\nu=0$}
  \end{subfigure}
  

\caption{Time dependencies of cluster size densities with active empty sites for $K(j,k)=j^{\mu}(1+k^{\nu})$}%
\end{figure}

Once again it can be seen that, in the large time, the time dependent solutions converge to corresponding cluster densities that were constructed recursively in the previous 
subsection. For more general exponents it is not possible to find
explicit solutions and hence one resorts to numerical computations. Also,
using moment analysis, one can extract important information on the large time behaviour. We discuss these in the next subsection.

\subsubsection{\textbf{General dynamics with active zero clusters}}

We consider kernels of the form $K(j,k)=j^{\mu}$ ($\mu\leq2)$ that allow transfer of monomers to zero clusters. To capture the change in the dynamics we look at the moments. 
When there is no external flux of particles the total mass of the system is conserved, i.e., $M_{1}(t)=M_{1}(0)$. 
Hence for any $0\leq\lambda<1$ the $\lambda-$moment is bounded $M_{\lambda}(t)<M_{1}(0).$ It is known from the general
theory that for $\mu\leq1$ the EDG system has a global solution and  all $\lambda-$moments for any finite interval is finite. 
So, we focus on the case $\mu>1$ which is more interesting. For $\lambda>\mu>1$ we first observe the following%
\begin{align*}
\dot{M}_{\lambda}(t)=\sum_{j=1}^{\infty}((j-1)^{\lambda}-j^{\lambda})j^{\mu
}c_{j}(t)M_{0}(t)\\
 +\sum_{j=0}^{\infty}((j+1)^{\lambda}-j^{\lambda}%
)c_{j}(t)M_{\mu}(t).
\end{align*}
For the summands we observe, by Taylor expansion%
\[
(j\pm1)^{\lambda}-j^{\lambda}=\pm\lambda j^{\lambda-1}%
+\frac{\lambda(\lambda-1)}{2}(j\pm\eta_{\pm})^{\lambda-2},%
\]
where $\eta_{\pm}$ are positive numbers less than one. Then, for $\lambda>1,$ we obtain the inequality%
\begin{align*}
\dot{M}_{\lambda}(t)\leq-\lambda M_{\lambda+\mu-1}(t)M_{0}(t) +CM_{\lambda+\mu-2}(t)M_{0}(t)\\
+CM_{\lambda-1}(t)M_{\mu}(t)+CM_{\lambda-2}(t)M_{\mu}(t).
\end{align*}
Assume that $1<\lambda\leq2$. Since $M_{0}(t),M_{\lambda-1}(t)\leq C$, using Jensen's inequality we get%
\[
\dot{M}_{\lambda}(t)\leq-\lambda M_{\lambda}^{1+\frac{\mu-1}{\mu}%
}(t)+CM_{\lambda}(t)
\]
which implies that $M_{\lambda}(t)<\infty.$ Now repeating the arguments for
arbitrary $2<\lambda\leq3$ and so on, we arrive at an important result that
$M_{\lambda}(t)$ is bounded for any $\lambda$. In fact we find numerically that for the kernel $K(j,k)=j^{\mu}$, the dynamics of EDG system with active zero clusters drives the size distribution to the equilibrium distribution.

Figure 4 shows the evolution of size distributions as a function of time for
the kernels with exponents $\mu=0,1,2$ and $\nu=0$, for monodisperse initial
conditions. One can easily check that the limit distributions for the
monodisperse initial conditions are exactly the self-consistent equilibrium
solutions with unit mass. It is remarkable that, even for exponents $\mu>1,$ 
the distributions and moments approach to equilibrium values where for the
EDG problem with passive zero clusters the solutions may not even exist \cite{esenturk_mathematical_2018}.

Another important question here is the rate of convergence to equilibria. From the exact solutions in Section III.B.1 one can explicitly see that the rate of approach of the cluster densities $c_j(t)$ to the equilibrium densities $c_j^e$ is exponentially fast, that is,
\[
|d_j(t)|=|c_j(t)-c_j^e|<Ce^{-rt}.
\]
More broadly, such exponential convergence was proved in \cite{esenturk_exchange-driven_2019} for a set of kernels that partly overlaps with the ones studied in this article, that is, $K(j,k)=ja(k)+b(j)$ where $a(k)$ is decreasing and $b(j)$ is increasing). Therefore one wonders if the same conclusion holds true here for general exponents. 

The plots below (Figure 5) suggest that the answer should be positive and the convergence should be exponential for $\mu \geq \nu>0$. However, for a general result, one needs to do a complete rigorous analysis of this claim  which we leave as a conjecture for future investigation.

\begin{figure}[h!]%
\centering

  \begin{subfigure}[b]{0.9\linewidth}
    \includegraphics[width=\linewidth]{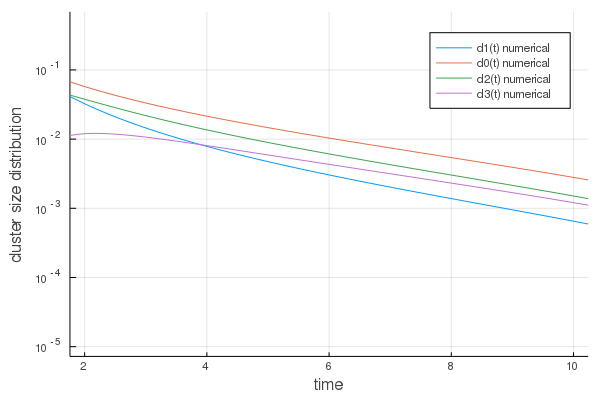}
    \caption{$\mu=0, \nu=0$}
  \end{subfigure}

  \begin{subfigure}[b]{0.9\linewidth}
    \includegraphics[width=\linewidth]{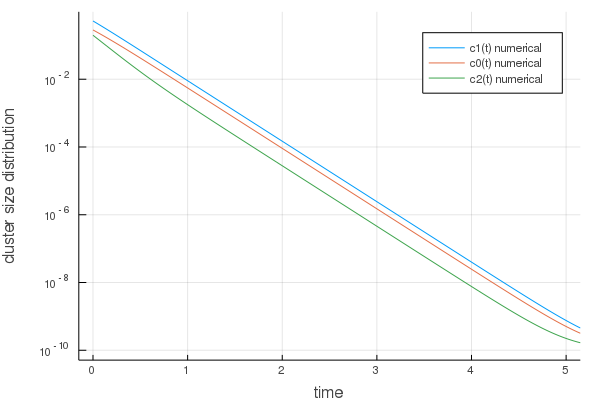}
    \caption{$\mu=2, \nu=0$}
  \end{subfigure}

\caption{Rate of approach of the cluster densities to equilibrium with active empty sites}%
\end{figure}

To complete the picture, we present the dynamics for $\mu=0, \nu=1$ for which there is no non-trivial equilibrium. Hence, the cluster densities, in this case, are not expected to settle on to non-zero values. Numerical computations show, instead, that the cluster densities decay in time just like in the passive zero-cluster case (see Figure 6).

\begin{figure}[h!]%
\centering

  \begin{subfigure}[b]{0.9\linewidth}
    \includegraphics[width=\linewidth]{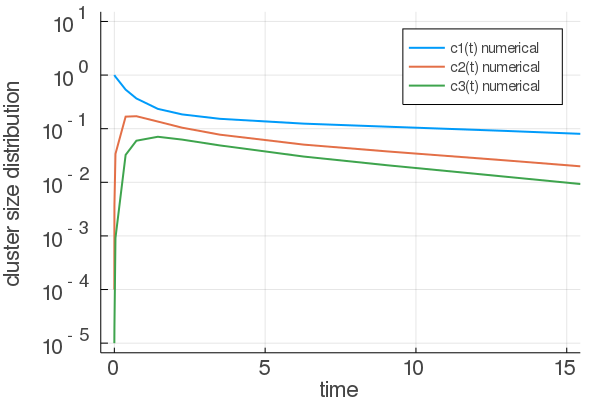}
    \caption{$\mu=0, \nu=1$}
  \end{subfigure}

\caption{Time dependence of cluster densities for a kernel which does not admit equilibrium ($\nu>\mu)$}%
\end{figure}

\section{Exchange-driven growth in a non-isolated system with source}

In this section we consider EDG in a non-isolated system with an external source that inputs  monomers at constant rate. As in the case of the isolated EDG
model, there is substantial difference in the behavior of solutions
with passive and active zero clusters. 

The EDG equation in the presence of monomer input is given by

\begin{equation}
\dot{c}_{0}=c_{1}\sum_{k=0}^{\infty}K(1,k)c_{k}-c_{0}\sum_{k=1}^{\infty
}K(k,0)c_{k}\text{ }\label{infode_so-0}%
\end{equation}

\begin{align}
\nonumber \text{ }\dot{c}_{j} &  =I\delta_{1,j}+c_{j+1}\sum_{k=0}^{\infty}K(j+1,k)c_{k}-c_{j}%
\sum_{k=0}^{\infty}K(j,k)c_{k}\\
&  -c_{j}\sum_{k=1}^{\infty}K(k,j)c_{k}+c_{j-1}\sum_{k=1}^{\infty
}K(k,j-1)c_{k}\text{ \ } \label{infode_so-j}%
\end{align}

\subsection{Passive zero clusters: Continuous growth and self-similarity}

As before we first consider the case of passive zero clusters with monomer input. The system was first studied theoretically in \cite{krapivsky_mass_2015} for symmetric kernels under a variety of settings. In particular, for the special case of the product kernel $K(j,k)=jk$ an exact solution was found. However, for more general product kernels analytical solutions are not available and we show the evolution of the cluster densities by numerical computations. 

We first present results for the size distribution of clusters with kernels ($K(j,k)=j^{\mu}k^{\nu}\theta(k))$. Figures 7(a),7(b) show the plot of cluster densities versus time for $c_{1},$ $c_{2},$ $c_{3}$ for exponents $\mu=1, \nu=0$ and $\mu=0, \nu=1$ with zero mass initial conditions. It is observed that cluster densities approach towards zero in the large time (despite the input of monomers). The decay is much slower for case (a) than case (b). This is probably due that, it is much quicker to export fresh mass to larger clusters with the kernel $K(j,k)=k$ than with kernels $K(j,k)=j\theta(k)$.

 For symmetric kernels, similar behaviour occurs, i.e., cluster densities approach to zero in the large time. Figure 7(c) shows numerical solutions for ($K(j,k)=j^{\mu}k^{\nu})$ with $\mu=\nu=1$ (also with zero mass initial condition). 

\begin{figure}[h!]%
\centering

 \begin{subfigure}[b]{0.9\linewidth}
    \includegraphics[width=\linewidth]{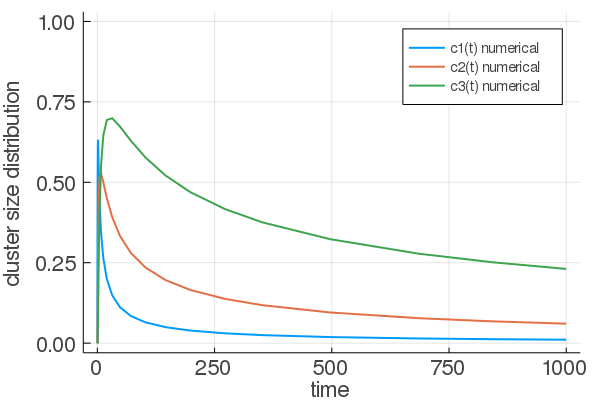}
    \caption{$\mu=1$, $\nu=0$}
  \end{subfigure}
   \begin{subfigure}[b]{0.9\linewidth}
    \includegraphics[width=\linewidth]{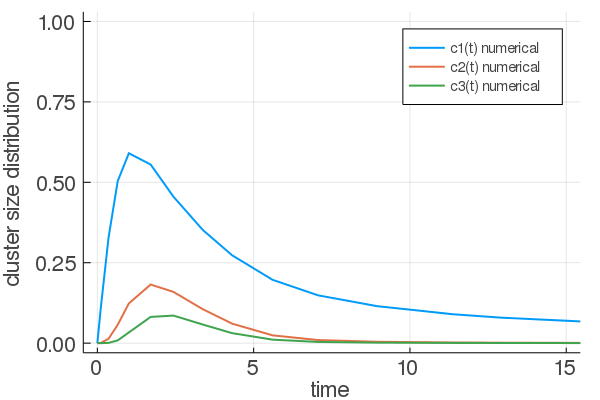}
    \caption{$\mu=0$, $\nu=1$}
  \end{subfigure}
  
  \begin{subfigure}[b]{0.9\linewidth}
    \includegraphics[width=\linewidth]{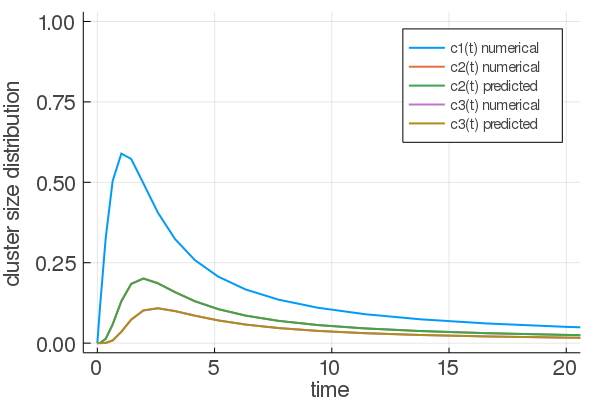}
    \caption{$\mu=\nu=1.0$}
  \end{subfigure}
  
\caption{Time dependencies of cluster size densities}%
\end{figure}

For symmetric kernels with non-integer values of $\mu,$ the approach that approximates
the infinite discrete system with a continuous equation is useful for the non-isolated system as well \cite{krapivsky_mass_2015}. If the kernels are
separable, one can neatly write an approximate equation for the system
replacing the difference terms in (\ref{infode-0})-(\ref{infIC}) with the
partial derivatives (keeping in mind of the monomer input). Then,
the solutions of the EDG after sufficiently long time can be approximated by
the solution of the following equation%

\begin{equation}
\partial_{t}c(t,x)=\partial_{xx}(x^{\mu}c(t,x))M_{\mu}(t)+I\delta
(x-1),\label{PDEor}%
\end{equation}
where we heuristically inserted $I\delta(x-1)$ term to account for the mass
source. On the right hand side of this equation, at large values of $x$, 
the source term drops out modifying only the nonlinear factor ($\lambda-$moment) in the first term. 
The $\lambda-$moment, in the continuum approximation, is given by%

\begin{equation}
M_{\lambda}(t)=\int_{0}^{\infty}x^{\lambda}c(t,x)dx.\label{mom-pde}%
\end{equation}

Due to the nonlinearity and nonlocality, one would not generally expect to find
analytic solutions to (\ref{PDEor}). We therefore look for special forms that will reduce the 
complexity. Consider the solutions in the scaling form

\[
c(t,x)\simeq\frac{1}{s(t)^{\alpha}}\phi(\frac{x}{s(t)}).
\]

Let $\xi=\frac{x}{s(t)}.$ In the new variables (\ref{mom-pde}) becomes

\[
M_{\lambda}(t)=s(t)^{1+\lambda-\alpha}\int_{0}^{\infty}\xi^{\lambda}f(\xi
)d\xi.
\]

The PDE (\ref{PDEor}) then reduces to%

\begin{equation}
-\frac{\alpha s^{\prime}(t)}{s(t)^{1+\alpha}}f(\xi)-\frac{s^{\prime}%
(t)}{s(t)^{1+\alpha}}\xi\frac{df(\xi)}{d\xi}=\frac{s(t)^{2\mu}}%
{s(t)^{1+2\alpha}}\frac{d^{2}(f(\xi)\xi^{\mu})}{d\xi^{2}}J(\xi),\label{SSequ}%
\end{equation}%
where $J(\xi)=\int_{0}^{\infty}\xi^{\mu}f(\xi)d\xi$.

\begin{figure}[h!]%
\centering

  \begin{subfigure}[b]{0.9\linewidth}
    \includegraphics[width=\linewidth]{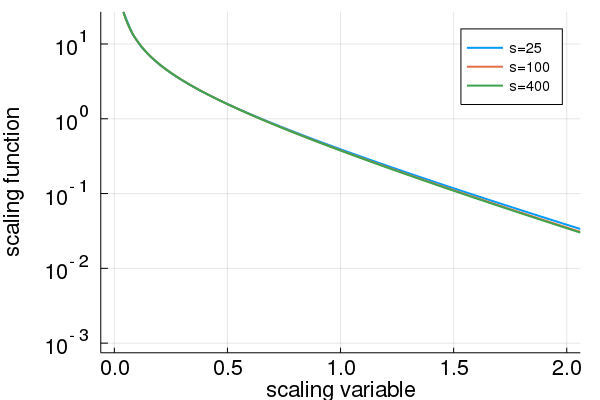}
    \caption{$\mu=1.0$}
  \end{subfigure}

 \begin{subfigure}[b]{0.9\linewidth}
    \includegraphics[width=\linewidth]{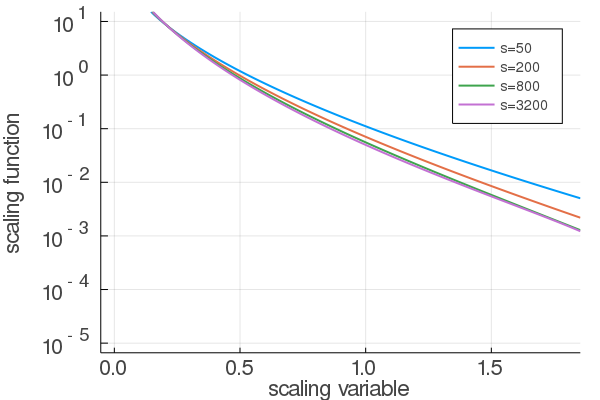}
    \caption{$\mu=1.2$}
  \end{subfigure}

\caption{Scaling function (log-scale) for the non-isolated EDG system with passive zero clusters}%
\end{figure}


We now separate the variables. Balancing the time dependent terms one gets
the relation $s^{\prime}(t)\sim s(t)^{2\mu-\alpha}$ which, upon integrating,
gives
\[
s(t)\sim t^{\frac{1}{1+\alpha-2\mu}}.
\]
Also, from the mass balance equation one has $M_{1}(t)=It$ which can be
written as%
\[
\left(  \int_{0}^{\infty}\xi f(\xi)d\xi\right)  s(t)^{2-\alpha}=It,
\]
from which we obtain%
\[
s(t)\sim t^{\frac{1}{2-\alpha}}.
\]
Combining the two expressions for the \textit{typical} mass $s(t)$ we find
$\alpha=\mu+1/2$ and hence
\[
s(t)=t^{\frac{2}{3-2\mu}}.
\]

The time dependence obtained for the \textit{typical} mass $s(t)$ indicates
faster growth for the non-isolated problem which is intuitively clear as the constant monomer input should speed up the growth. Once again we can
see, in Figure 8, that numerical computations agree very well with the the theoretical
predictions, that is, the parametric curves $(\frac{j}%
{s(t)},s(t)^{\alpha}c_{j}(t))$ collapse into a single curve demonstrating self-similarity. The collapse is more visible for large time and large cluster sizes. The delay in the collapse for $\mu=1.2$ can be attributed to the size of the simulation being finite ($N=7000$). As the exponents approach the critical value for gelling ($\mu_{c}=1.5$ one needs larger and larger simulation sizes in order to capture the onset of self-similarity. 

\subsection{Dynamics with Active Zero Clusters and Linear (in time) Growth}

In this section of this paper we again consider the kernels $K(j,k)=\frac{1}{2}\,j^{\mu}(1+k^{\nu})$ which allow hopping to zero clusters. 
For the isolated system it was shown that the system approaches to a non-trivial equilibrium when $\nu < \mu$. Hence it is interesting to know how input of particles affects the dynamics. The coupled EDG equations read%

\begin{equation}
\dot{c}_{0}=\frac{1}{2}c_{1}(M_{0}+M_{\nu })-\frac{1}{2}(1+\delta
_{v,0})c_{0}M_{\mu },  \label{EDGS0}
\end{equation}

\begin{equation}
\begin{split}
\text{ }\dot{c}_{j}& =\frac{1}{2}(j+1)^{\mu }c_{j+1}(M_{0}+M_{\nu })-\frac{1%
}{2}j^{\mu }c_{j}(M_{0}+M_{\nu }) \\
& -\frac{1}{2}(1+j^{\nu })c_{j}M_{\mu }+\frac{1}{2}(1+(j-1)^{\nu
})c_{j-1}M_{\mu }+I\delta _{j,1}.
\end{split}  \label{EDGS}%
\end{equation}	
It is difficult to solve
this system exactly for arbitrary $\mu ,\nu .$ But one can obtain
approximate analytic solutions in the following way. We first consider the
case $\mu =1,$ $\nu =0$ (i.e., $K(j,k)=j).$ The coupled system simplifies
since $M_{0}(t)=It$ and $M_{1}(t)=It.$ We can then divide all terms in (\ref%
{EDGS0}) - (\ref{EDGS}) equations by $It$ and absorb the coefficient $\frac{1%
}{It}$ by defining a new time variable $\tau (t)=t^{2}/2.$ For large $\tau $
the coefficient of the source term on the right hand side will be
insignificant (decaying as $\sim \frac{1}{\sqrt{\tau }}).$ Therefore, the
approximate solutions of equations (\ref{EDGS0}) - (\ref{EDGS})
will be nothing but the solutions of the equilibrium problem obtained in
Eq.~(\ref{eq-eqm}) in Section III.B. Going back to the original time variable
one finds that the real time approximate solutions of (\ref{EDGS0}) - (\ref%
{EDGS}) are the equilibrium solutions of the problem without a source
(isolated system) with modified mass and volume as $M_{1}=M_{0}=It$, that is,%
\begin{equation}
c_{j}(t)=\frac{e^{-1}}{j!}It\text{ }~\ \ \{\text{for }\mu
=1, \nu=0\}.\label{dyn-equi-mu1}%
\end{equation}

\begin{figure}[h!]%
\centering

  \begin{subfigure}[b]{0.9\linewidth}
    \includegraphics[width=\linewidth]{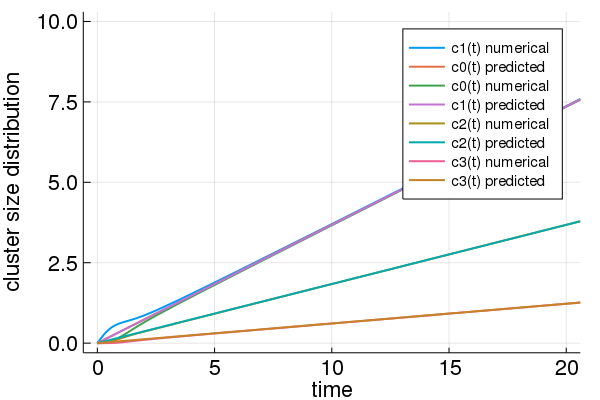}
    \caption{$\mu=1, \nu=0$}
  \end{subfigure}

 \begin{subfigure}[b]{0.9\linewidth}
    \includegraphics[width=\linewidth]{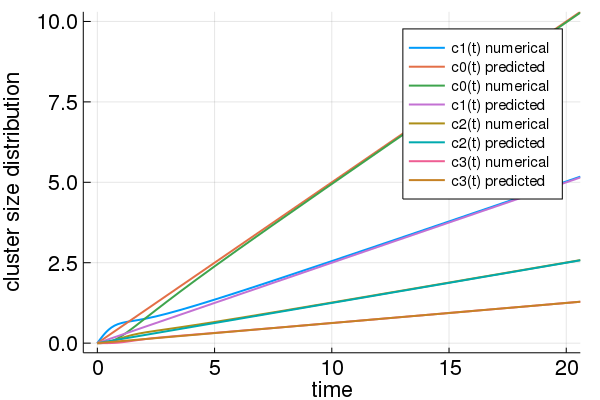}
    \caption{$\mu=0, \nu=0$}
  \end{subfigure}

  \begin{subfigure}[b]{0.9\linewidth}
    \includegraphics[width=\linewidth]{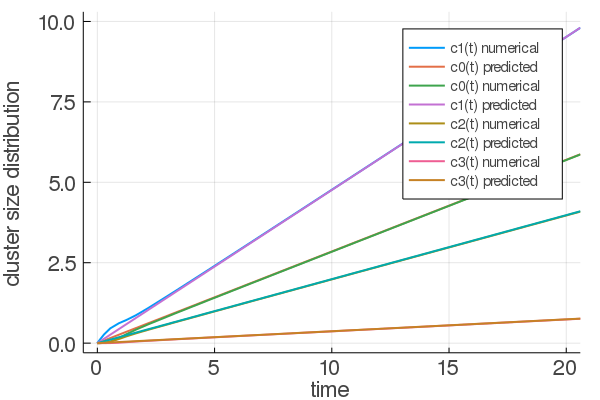}
    \caption{$\mu=2, \nu=0$}
  \end{subfigure}

\caption{Time dependence of cluster densities for the non-isolated EDG system with active zero clusters}%
\end{figure}

Using this ansatz for the general product kernel $K(j,k)=\frac{1}{2}j^{\mu}(1+k^{\nu})$, we infer that, $c_{j}(t)=tc_j^e$ is the approximate solution of
(\ref{EDGS0}) - (\ref{EDGS}) where $c_j^e$ is the self-consistent equilibrium solution
(\ref{eqm-separable}) obtained in Section III.B. Another argument is that, for the isolated system, the total mass and total volume (number of clusters) alone determine the eventual equilibrium distribution for which the time dependent solutions approach exponentially fast.
For the non-isolated system under consideration, since the external forcing is linear in time, asymptotically, the system quickly relaxes to the "equilibrium distribution" given by the current total mass and total volume of the system.

To test the argument, we also find approximate analytic expressions for the exponents $\mu=\nu=0$ and $\mu=\nu=1$. Extracting the expression for the
self-consistent equilibrium solutions for these exponents one has%

\[
c_{j}(t)=\frac{It}{2^{j+1}}\text{ }\{\text{for }\mu=0\}.
\]
Similarly for $\mu=2$, $\nu=0$, we expect the solution are given by $c_{j}(t)=tc_j^e$ where $c_j^e$ is the corresponding equilibrium solution from Section III. The results are in excellent agreement (asymptotically) with the numerical solutions  as shown
in Figure 9. While this regime is new in the context of EDG, very similar behaviour has recently been observed in a growth model with aggregation and fragmentation in the presence of an external source of particles \cite{bodrova2019kinetic}. This suggests that there is a novel underlying mechanism at work that is not specific to this particular growth model.

Finally, for the exponents $\mu=0$, $\nu=1$ we know from Section III that there are no non-trivial equilibrium solutions. Hence there is no state to relax onto and therefore the growth, i.e., disappearance of finite clusters, continues. The radical deviation from the linearly growing cluster densities to decay of clusters can also seen from the numerical results as shown in Figure 10 below.

\begin{figure}[h!]%
\centering

  \begin{subfigure}[b]{0.9\linewidth}
    \includegraphics[width=\linewidth]{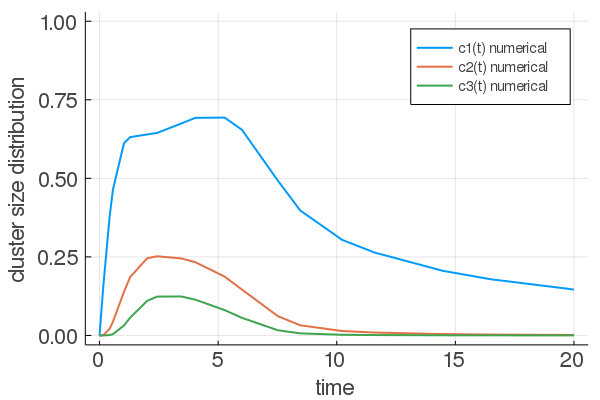}
    \caption{$\mu=0, \nu=1$}
  \end{subfigure}

\caption{Time dependence of cluster densities for the non-isolated EDG system with active zero clusters}%
\end{figure}

\section{CONCLUSION}

In this article, we studied the exchange-driven growth model for two classes
of dynamics with (i) the passive zero clusters where interactions take place only between
non-zero clusters, (ii) the active zero clusters that allow hopping of
particles from non-zero clusters to zero clusters. For each case,
we studied the dynamics for the isolated (purely exchange interaction) and non-isolated
system (with input of monomers). Our key finding is that the seemingly small change from passive to active zero clusters alters the dynamics fundamentally.

For the isolated system we found that for a system with active zero clusters
the classical indefinite growth trend \cite{ben-naim_exchange-driven_2003} is broken and the cluster densities can approach to non-trivial equilibrium values provided the kernel does not grow too fast as a function of the size of the arrival cluster. For a broad set of kernels we showed these using theoretical arguments and numerical computations. The
numerical results are in excellent agreement with the theoretical ones.

For the non-isolated system, the inclusion of zero clusters into the dynamics again substantially changes the behaviour of solutions. In systems with passive zero clusters, for product type kernels \cite{krapivsky_mass_2015}, we numerically and theoretically
demonstrated that the cluster densities approach to zero (indefinite
growth)\ and the size distribution of clusters take self similar form in the large time. On the other hand if interaction with the zero clusters is allowed (active case), then an
interesting behaviour occurs: the inputted mass gets distributed evenly among all clusters and cluster densities grow linearly (asymptotically) in time.

\section{APPENDICES}



\subsection{Derivation for time dependence of cluster size densities in the passive case for $\mu=0$ and $\nu=1$}

\bigskip We wish to solve the EDG system (\ref{infode-0}) - (\ref{infIC})
with the initial conditions $c_{j}(0)=\delta _{j,1}.$ We seek solutions of
the form 
\begin{equation}
c_{j}(t)=A(t)a(t)^{j-1}.  \label{cj-form}
\end{equation}%
where the initial conditions require that $A(0)=1$ and $a(0)=0.$ By the \
conservation of mass $\sum_{j\geq 1}jc_{j}(t)=1,$ plugging the form (\ref%
{cj-form}) in the summation, one finds $A(t)$ and $a(t)$ are related by%
\begin{equation*}
A(t)=(1-a(t))^{2}.
\end{equation*}

Now, we put (\ref{cj-form}) back in the EDG system and look for $a(t)$ which
satisfies ODE system. Taking the time derivative of (\ref{cj-form}) the left
hand side of (\ref{infIC}) can be written as%
\begin{equation}
\dot{c}_{j}=2(1-a)(-\dot{a})a^{j-1}+(1-a)^{2}(j-1)a^{j-2},
\label{cj_der-lhs-app}
\end{equation}%
while the right hand, by the seperability of the kernel, reduces to%
\begin{equation}
\dot{c}_{j}=c_{j+1}-c_{j}-(jc_{j}-(j-1)c_{j-1})N_{0}  \label{cj_der-rhs-app}
\end{equation}%
where $N_{0}(t)=\sum_{j\geq 1}c_{j}(t)$ is the number of clusters. Using the
form for the $c_{j}$ it is easy to see that 
\begin{equation*}
N_{0}(t)=(1-a(t))^{2}\sum_{j\geq 1}a(t)^{j-1}=1-a(t)
\end{equation*}%
Using this back (\ref{cj_der-rhs-app}) one gets%
\begin{equation*}
RHS=(1-a)^{2}a^{j-2}(a^{2}-a-(1-a)ja+(1-a)(j-1))
\end{equation*}%
Equalizing both expression for $c_{j}$ in (\ref{cj_der-lhs-app}) and (\ref%
{cj_der-rhs-app}) the terms we see that $(1-a)a^{j-2}$ are common and cancel
out. We are the left with%
\begin{equation*}
2(-\dot{a})a+(1-a)(j-1)=(1-a)^{2}(-2a+(1-a)(j-1))
\end{equation*}%
In order this equality to hold the function $a(t)$ should satisfy the simple
ODE%
\begin{equation*}
\dot{a}=(1-a)^{2}.
\end{equation*}%
Upon solving with the initial condition we find 
\begin{equation*}
a(t)=\frac{t}{1+t}
\end{equation*}%
which yields, for $c_{j},$%
\begin{equation*}
c_{j}(t)=\left( \frac{t}{1+t}\right) ^{j-1}\frac{1}{(1+t)^{2}}.
\end{equation*}

\subsection{Time dependence of cluster size densities in the active case for $\mu=1$ and $\nu=0$}

Define the generating function%
\begin{displaymath}
G(t,z) = \sum_{k=0}^\infty c_k(t)\,z^k.
\end{displaymath}
Using Eqs. (\ref{infode-0})-(\ref{infIC}), the generating function satisfies
\[
\partial_{t}G(t,z)=(1-z)\partial_{z}G(t,z)M_{0}-M_{1}(1-z)G(t,z).
\]
The characteristic equations are given by%
\[
\frac{dt}{ds}=1,\text{ \ }\frac{dz}{ds}=-(1-z)M_{0},\text{ \ \ }\frac{dG}%
{ds}=-M_{1}(1-z)G.
\]
From these equations we find $z(s)$ as%
\[
1-z(s)=(1-z_{0})e^{M_{0}s}.
\]
The equation for $G$ can be solved to give%
\begin{align*}
G(t,z) &  =G_{0}(z_{0})\exp(-(1-z)(1-e^{-M_{0}t})\frac{M_{1}}{M_{0}})\\
&  =G_{0}((1-z)e^{-M_{0}t})\exp(-(1-z)(1-e^{-M_{0}t})\frac{M_{1}}{M_{0}}).
\end{align*}
Let $g(t)=(1-e^{-M_{0}t})M_{1}/M_{0}.$ If we assume monodisperse initial
condition, i.e., $G_{0}(z)=(M_{0}-M_{1})+M_{1}z$ then, for the cluster
distributions, we can write $G(t,z)$ as a series%
\begin{align*}
G(t,z) &  =(M_{0}-M_{1}e^{-M_{0}t})e^{-g(t)+g(t)z}+zM_{1}e^{-M_{0}%
t-g(t)+g(t)z}\\
&  =(M_{0}-M_{1}e^{-M_{0}t})e^{-g(t)}\sum_{n=0}^{\infty}\frac{g(t)^{n}}%
{n!}z^{n}\\
& +M_{1}e^{-M_{0}t-g(t)}\sum_{n=1}^{\infty}\frac{g(t)^{n-1}}{(n-1)!}z^{n}
\end{align*}
and extract the coefficients of the power series to give%
\[
c_{j}(t)=(M_{0}-M_{1}e^{-M_{0}t})e^{-g(t)}\frac{g(t)^{j}}{j!}+e^{-M_{0}%
t-g(t)}\frac{g(t)^{j-1}}{(j-1)!}.
\]

\textbf{Acknowledgements. } We thank Stefan Grosskinsky for drawing our attention to the importance of the active case and for helpful discussions. EE is supported by Marie Sklodowska-Curie Individual Fellowship H2020-MSCA-IF-2015 project no. 705033.

%

\end{document}